\documentclass[11pt,reqno,a4paper]{amsart}
\pdfoutput=1
\usepackage{amsaddr}     
\usepackage{amssymb}

\usepackage{enumitem}

\usepackage{fancyhdr}
\usepackage{graphicx}
\usepackage[norelsize]{algorithm2e}
\SetAlCapSkip{1em}

\newcommand{\bx}{{\boldsymbol{x}}}
\newcommand{\bX}{{\boldsymbol{X}}}
\newcommand{\by}{{\boldsymbol{y}}}
\newcommand{\bY}{{\boldsymbol{Y}}}
\newcommand{\bn}{{\boldsymbol{n}}}
\newcommand{\bH}{{\boldsymbol{H}}}
\newcommand{\bW}{{\boldsymbol{W}}}

\newcommand{\bP}{{\boldsymbol{P}}}
\newcommand{\bT}{{\boldsymbol{T}}}
\newcommand{\bV}{{\boldsymbol{V}}}
\newcommand{\bS}{{\boldsymbol{S}}}
\newcommand{\bI}{{\boldsymbol{I}}}
\newcommand{\bb}{{\boldsymbol{b}}}
\newcommand{\bQ}{{\boldsymbol{Q}}}
\newcommand{\bC}{{\boldsymbol{C}}}
\newcommand{\bR}{{\boldsymbol{R}}}
\newcommand{\bB}{{\boldsymbol{B}}}

\newcommand{\bGamma}{{\boldsymbol{\Gamma}}}

\usepackage[round]{natbib}

\begin{document}

\title[Multi-frequency image reconstruction]
{{\tiny \textsf{SKA Pathfinders Radio Continuum Surveys 2015 - SPARCS}}\bigskip\\
Multi-frequency image reconstruction\\for radio interferometry.\\
A regularized inverse problem approach.}

\author[A. Ferrari, J. Deguignet, C. Ferrari, D. Mary, A. Schutz]
{A. Ferrari, J. Deguignet, C. Ferrari, D. Mary, A. Schutz}
\address{Lab. Joseph-Louis Lagrange,\\
Universit\'e  de Nice Sophia Antipolis, CNRS, \\
Observatoire de la C\^ote d'Azur,  Nice, France}
\author[O. Smirnov]{O. Smirnov }
\address{Centre for Radio Astronomy Techniques \& Technologies (RATT),\\
Department of Physics and Electronics, Rhodes University,\\ South Africa,
and SKA South Africa.}

\thanks{This work was partly supported by the Agence Nationale pour la Recherche, France : MAGELLAN project, ANR-14-CE23-0004-01.
This work was granted access to the HPC and visualization resources of ``Centre de Calcul Interactif'' hosted by ``Universit\'e Nice Sophia-Antipolis''}

\begin{abstract}
We describe a ``spatio-spectral'' deconvolution algorithm for wide-band imaging in
radio interferometry. In contrast with the existing multi-frequency reconstruction algorithms,
the proposed method does not rely on a model of the sky-brightness spectral 
distribution. This non-parametric approach can be of particular interest for
the new generation of low frequency radiotelescopes.
The proposed solution formalizes the reconstruction problem as
a convex optimization problem with spatial and spectral regularizations.
The efficiency of this approach has been already proven for narrow-band image reconstruction and the present  contribution can be considered as its extension to 
the multi-frequency case. Because the number of frequency bands multiplies the size 
of the inverse problem, particular attention is devoted to the derivation
of an iterative large scale optimization algorithm. 
It is shown that the main computational bottleneck of the approach, which lies in
the resolution of a linear  system, can be efficiently overcome by a fully
parallel implementation  w.r.t. the 
frequencies, where each processor reconstructs a narrow-band image.
All the other  optimization steps are extremely fast. 
A parallel implementation of the algorithm in Julia is publicly available at 
\texttt{https://github.com/andferrari}.
Preliminary simulations illustrate the performances of the method and its
ability to reconstruct complex spatio-spectral structures.
\end{abstract}

\maketitle

\section{Introduction}

Recently, much attention has been paid to the development of image reconstruction
algorithms for the incoming and future radio facilities.  Most recent contributions, 
such as \citep{sara,moresane,purify} or \citep{Garsden15}
heavily rely on  sparse estimation.  In the wake of 
CLEAN algorithm \citep{clean}  and its multiresolution variants, 
sparse models have indeed  proved  in the last decades to be a powerfull approach
for radiointerferometric image reconstruction in particular, and
for the resolution of inverse problems in general. 

In addition to their high spatial resolution, the wide bandwidths of the
new generation of radio  interferometers  makes possible the reconstruction
of complex  spectral structures. 
The Square Kilometre Array  (SKA) and its  precursors will achieve (sub-)arcsec resolution over hundreds of MHz instantaneous bandwidths and with a tremendously broad band coverage (see Table 1 in the SKA1 System Baseline Design document \citep{Dewdney13}).
The reconstruction of both  spatial  and spectral behaviour of continuum radio sources is an essential tool to characterize the astrophysical origin of their detected radiation, e.g. \citep{krausb}. \citep{RauMSMF} opened the way to
multi-frequency deconvolution algorithms, which aim to reconstruct simultaneously
spatial and spectral structures. The approach proposed in \citep{RauMSMF}
relies on the  parameterization of the frequency-dependent brightness distribution
as a power law with a varying index. A Taylor expansion is adopted to model the  flux   dependence in frequency of astrophysical radio sources, whose synchrotron or thermal  spectra can be described by power-laws \citep{conway90}.
From an estimation point of view,
the ratio  of  the additional unknowns  
introduced by   such a multi-frequency model (e.g. the average spectral-indexes and the spectral-curvatures
for a second order model)   to  the additional equations (the multiplication of measurements by the number of
frequency bands) is clearly in favor of a multi-frequency reconstruction approach.
Explicit models of the frequency
 dependence of radio sources have also been introduced and exploited in \citep{Junklewitz14}  
and \citep{Bajkova11}.  In  \citep{Junklewitz14},   the authors 
propose to   address the estimation problem using a Bayesian framework.   
\citep{Bajkova11} proposes to   constrain a
maximum entropy estimation algorithm in order to account for the frequency dependence of the intensities. 

These ``semi-parametric'' methods rely on spectral models and thus clearly offer advantages and estimation accuracy when the model is indeed appropriate. Across the broad frequency coverage of current radio facilities however, radio sources exhibiting complex spectral shapes (not simple power laws) can be expected. Such sources can show one or more relative minima, breaks and turnovers \citep{Kellerman74}. More recently,
\citep{Scaife12} have  shown that second order broadband spectral models are often insufficient for the new generation of low frequency telescopes such as the Low Frequency Array (LOFAR). A non-parametric approach in the multi-frequency reconstruction of radio sources is 
also definitely needed for the full Stokes (Q, U, V) wide-band imaging, where Taylor expansion is not the appropriate physical model to be adopted \citep{RauMSMF}. 
In the same direction, \citep{Wenger:2014ty} have proposed to relax the spectral power-law model
by formulating the problem as an inverse problem with a smooth spectral regularization allowing for local deviations.

This communication proposes to fully formalize  multi-frequency image reconstruction 
as a regularized inverse problem. As such, it extends the models proposed in 
\citep{sara,purify,Garsden15} to the multi-frequency
observation mode by adding a spectral regularization term. It is worthy to note that
the proposed approach shares formal similarities with a spatio-spectral image reconstruction algorithm
recently proposed for optical interferometry \citep{PAINTER,admmthiebaut}.

Section 2. introduces the data model and the  inverse problem criterion. 
Section 3. derives an efficient optimization algorithm to minimize the related convex problem.
The time consuming steps of this iterative algorithm can be 
parallelized frequency-wise
leading to an overall computation time comparable
to a narrow-band reconstruction algorithm. Section 3. presents preliminary
simulation results where the observations are obtained using the
MeqTrees package~\citep{meqtrees}.

\section{Models and notations}

We denote by $\bx_\ell$ the column vector collecting  the sky intensity image at 
observing frequency  $\nu_\ell$, $\ell=1,\ldots,L$. 
The sky image map $\bx_\ell$ is related to the ``dirty image'' at frequency 
$\nu_\ell$, denoted as $\by_\ell$, by
\begin{equation}\label{nbmodel}
\by_\ell = \bH_\ell \bx_\ell + \bn_\ell
\end{equation}
where $\bn_\ell$ is a noise vector. $\bH_\ell$ is a convolution matrix  containing 
the single-frequency point-spread function (PSF), which
 includes  weighting factors and visibility mapping on a spatial frequency grid \citep{RauMSMF}.
Stacking the images at different frequencies in a single vector leads to the wideband
model
\begin{equation} \label{wbmodel}
\by = \bH \bx + \bn
\end{equation}
where $\bH$ is a block diagonal matrix containing the $\bH_\ell$ matrices on the main diagonal.

Note  that we consider herein  without loss of generality the model connecting the 
sky intensity to the dirty map. The same algorithm  as the one described below can be derived
from the model connecting the sky intensity to the complex visibilities by replacing in (\ref{nbmodel})   $\by_\ell $
 by the sample visibilities and $\bH_\ell$ by the operator mapping
 the sky intensity to the visibilities at the observing frequency  $\nu_\ell$, see Appendix A.2 of  \citep{RauMSMF}.

Eq.~\ref{wbmodel} defines an ill-posed linear inverse problem.
Among all existing  methods  to solve inverse problems, we
will focus on sparse regularization and optimization techniques, 
which have gained much popularity during the last decade.
The regularization game consists in minimizing a criterion composed by a 
fidelity term and a regularization term $f_\text{reg}$ linked to some prior on the solution.  We will consider here
the objective:
\begin{equation} \label{costreg}
\min_\bx\; \frac{1}{2}  \|\by - \bH \bx \|^2 + f_\text{reg}(\bx)
\end{equation}

Many recent works have shown that regularization based on  sparse 
representations in  appropriate transform domain(s) of the  intensity map being restored
can be very effective. This approach can be formulated in an analysis
framework: the criterion is minimized directly w.r.t. the image parameters, i.e., $\bx$. An alternative formulation,
not considered in the present study,
is the synthesis framework: the criterion is then minimized w.r.t. the coefficients
of the image decomposition. Those can be much more numerous than the image parameters if the decomposition is redundant - and it should to be efficient. This leads to an increased computational load w.r.t. the analysis framework. 
These two formalisms are discussed in \citep{elad2007}. In practice and performance-wise, it is still unclear which approach should be preferred and under which conditions.
The efficiency of both approaches for narrow-band radio-interferometric imaging has been 
proven in \cite{sara,purify} (analysis), \cite{moresane} (analysis-synthesis), \cite{Garsden15} (synthesis). Similarly to the first  two references, we will focus on
a sparse analysis prior and in complement to the  classical positivity and quadratic smoothing regularization, 
we will consider a regularization of the form:
\begin{equation}
f^\text{reg}(\bx) = \boldsymbol{1}_{\mathbb{R}^+}(\bX) +\frac{\mu_\varepsilon}{2} \| \bX \|_\text{F}^2 	+ \mu_s \| \bW_s \bX\|_1	  + \mu_\nu \| \bX \bW_\nu \|_1	 
  \label{eq:RegAll}
\end{equation}
where $\bX$ is the matrix defined as $\bX = (\bx_1,\ldots,\bx_{l_\nu})$ and
$\bW_s$ and $\bW_\nu$ are the matrices associated  with respectively the spatial and spectral 
analysis regularizations. In practice these matrices take the form of (usually redundant) dictionaries with dedicated fast transforms, some examples will be given below.  It is important to  underline the central role of the 
last regularization term with parameter $\mu_\nu$ in Eq.~\ref{eq:RegAll}. This term prevents the  optimization problem (\ref{costreg}) to be separable w.r.t. the $\bx_\ell$. This makes the sparse spatial and spectral priors imbedded and the regularization truly spatio-spectral.

\section{Optimization algorithm}

We propose to find the solution of the convex problem  (\ref{costreg},\ref{eq:RegAll})
using the Alternating Direction Method of Multipliers (ADMM) algorithm.
For a recent comprehensive review of ADMM see \citep{Boyd10}.
Convergence of ADMM was demonstrated in \citep{EcksteinBertsekas}. As it will be shown below, this method is particularly
interesting to solve large-scale problems such as  (\ref{costreg},\ref{eq:RegAll}), as it leads to successive steps
that can be parallelized w.r.t. the images at each frequency (i.e., the  $\bx_\ell$, columns of $\bX$) or w.r.t. 
the  spectra at each pixel position  (i.e., the rows of $\bX$).

The end of this section is devoted to the derivation of this algorithm. 
The optimization problem (\ref{costreg},\ref{eq:RegAll}) is equivalent to:
\begin{align}
&\min_ \bX\; \frac{1}{2}  \|\bY - \bH \bX \|^2 + 
\boldsymbol{1}_{\mathbb{R}^+}(\bP) +\frac{\mu_\varepsilon}{2} \| \bX \|_\text{F}^2 	
+ \mu_s \| \bT\|_1	  + \mu_\nu \| \bV \|_1	\label{admm1}\\
&\text{subject to: }\bP =\bX,\; \bT = \bW_s \bX,\; \bV = \bS \bW_\nu,\; \bS =\bX 	\label{admm2}
\end{align}
where   $\bY = (\by_1,\ldots,\by_{l_\nu})$. Auxiliary variables $\bP$, $\bT$, $\bV$ and $\bS$ have  
associated
 Lagrange multipliers
$\bGamma^P$, $\bGamma^T$, $\bGamma^V$ and $\bGamma^S$ and augmented Lagrangian  parameters
$\rho_P$, $\rho_T$, $\rho_V$ and $\rho_S$.
Denoting with a $+$ superscript the updated quantities, the alternated minimizations 
of the augmented Lagrangian of (\ref{admm1},\ref{admm2}) give:
\begin{enumerate}[label=\Roman*.]
\item \emph{Minimization w.r.t. $\bX$.} 
This   step operates separately on  each  frequency image $ \bx_\ell$.
It requires to solve for each frequency $\nu_\ell$ the linear system 
\begin{equation}
\bQ_\ell \bx_\ell^+ = \bb_\ell,\; \ell=1\ldots,l_\nu \label{linsystemX}
\end{equation}
where:
\begin{align}
&\bQ_\ell = \bH_\ell^\top\bH_\ell +\rho_T \bW_s^\top \bW_s+(\mu_\epsilon +\rho_S+\rho_P)\bI \label{matQ}\\
& \bB = \bH^\top \bY + \bW_ s^\top(\bGamma^T + \rho_T\bT)
+ \bGamma^P +\rho_P\bP +   \bGamma^S +\rho_S\bS \label{matB}
\end{align}
and $\bb_\ell$ are the columns of $\bB$.

\item \emph{Minimization w.r.t. $\bP$.} 
Defining $\tilde{\bP} = \bX - \rho_P^{-1}\bGamma_P$, this minimization
simplifies to the proximity operator:
\begin{equation}
\min_\bP \; \boldsymbol{1} _{\mathbb{R}^+}(\bP) + \frac{\rho_P}{2}\|\bP -\tilde{\bP} \|_F^2
\end{equation}
and leads to the positive projection of each element $\tilde{\bP}_{i,j}$:
\begin{equation}
\bP_{i,j}^+ = \max(0,\tilde{\bP}_{i,j})
\end{equation}

\item \emph{Minimization w.r.t. $\bT$.} 
Defining $\tilde{\bT} = \bW_s\bX -\rho_T^{-1} \bGamma_T$, minimisation w.r.t. $\bT$
simplifies to the proximity operator:
\begin{equation}
\min_\bT\; \mu_s \|\bT\|_1 + \frac{\rho_T}{2} \| \bT - \tilde{\bT} \|_F^2 
\end{equation}

Consequently, each element $\bT_{i,j}$ of $\bT$ is updated according to the soft thresholding operator:
\begin{equation}
\bT_{i,j}^+ = \tilde{\bT}_{i,j}\max\left(0, 1 - \frac{\rho_T^{-1}\mu_s}{|\tilde{\bT}_{i,j}|}\right) 
\end{equation}

\item \emph{Minimization w.r.t. $\bV$.} 
Defining $\tilde{\bV} = \bS\bW_\nu -\rho_V^{-1} \bGamma_V$, minimisation w.r.t. $\bV$
simplifies to:
\begin{equation}
\min_\bV \;\mu_\nu \|\bV\|_1 + \frac{\rho_V}{2} \| \bV - \tilde{\bV} \|_F^2 
\end{equation}

Each element $\bV_{i,j}$ of $\bV$ is updated according to the soft thresholding operator:
\begin{equation}
\bV_{i,j}^+ = \tilde{\bV}_{i,j}\max\left(0, 1 - \frac{\rho_V^{-1}\mu_\nu}{|\tilde{\bV}_{i,j}|}\right)
\end{equation}

\item \emph{Minimization w.r.t. $\bS$.} 
This steps operates separately on each voxel.
It requires to solve the linear system $\bS^+ \bR = \bC$
where:
\begin{align}
&\bR = \rho_V \bW_\nu\bW_\nu^\top + \rho_S \bI\\ 
& \bC = (\bGamma_V+\rho_V\bV)\bW_\nu^\top - \bGamma_S +\rho_S\bX
\end{align}
This step is the twin of step I. (minimization w.r.t. $\bX$). 
Choosing for $\bW_\nu$ an orthonormal transform, or a union of $m_\nu$ orthonormal transforms, is of particular interest as it leads to:
\begin{equation}
\bS^+ = \frac{1}{m_\nu\rho_V  + \rho_S}
\left((\bGamma_V+\rho_V\bV)\bW_\nu^\top - \bGamma_S +\rho_S\bX\right)
\end{equation}

\item \emph{Update of the Lagrangian multipliers.} Finally, the Lagrangian multipliers
are updated according to the standard way:
\begin{align}
& \bGamma_P^+ = \bGamma_P +\rho_P(\bP-\bX) \\
& \bGamma_T^+ = \bGamma_T +\rho_T(\bT-\bW_s\bX) \\
& \bGamma_V^+ = \bGamma_V +\rho_V(\bV-\bS\bW_\nu) \\
& \bGamma_S^+ = \bGamma_S +\rho_S(\bS-\bX) 
\end{align}
\end{enumerate}
The previous six steps are iterated until convergence
(see \citep{Boyd10} for the stopping criterion). This leads to
algorithm \ref{algo:ADMMalgo}, where $\text{S}_{r}(\cdot)$
denotes the element-wise soft thresholding operator with threshold $r$.

\begin{algorithm}[h]
Initialize to zero $\bX$, $\bP$, $\bT$, $\bV$, $\bS$, $\bGamma_P$, 
$\bGamma_T$, $\bGamma_V$ and $\bGamma_S$\;
\DontPrintSemicolon 
\Repeat{stopping criterion is satisfied.}{
  \tcc{update primal variables}
  \ForPar{$\ell=1\ldots,l_\nu$}{
    Compute $\bb_\ell$  using (\ref{matB})\;    
	\tcc{conjugate gradient algorithm}
  	Solve $\bQ_\ell \bx_\ell^+ = \bb_\ell$ where $\bQ_\ell$ is given by (\ref{matQ})}
  $\bP \gets \max(\boldsymbol{0},\bX - \rho_P^{-1}\bGamma_P)$\;
  $\bT \gets \text{S}_{ \mu_s/\rho_T}(\bW_s\bX -\rho_T^{-1} \bGamma_T)$\;
  $\bV \gets \text{S}_{\mu_\nu/\rho_V} (\bS\bW_\nu -\rho_V^{-1} \bGamma_V)$\;
  $\bS \gets (m_\nu\rho_V  + \rho_S)^{-1}\left((\bGamma_V+\rho_V\bV)\bW_\nu^\top - \bGamma_S +\rho_S\bX\right)$\;
  \tcc{update dual variables}
  $\bGamma_P \gets \bGamma_P +\rho_P(\bP-\bX)$\;
  $\bGamma_T \gets \bGamma_T +\rho_T(\bT-\bW_s\bX)$\;
  $\bGamma_V \gets \bGamma_V +\rho_V(\bV-\bS\bW_\nu)$\;
  $\bGamma_S \gets \bGamma_S +\rho_S(\bS-\bX)$\; 
}
\Return{$\bX$}
\caption{Multi-frequency reconstruction alorithm.}
\label{algo:ADMMalgo}
\end{algorithm}


Similarly to the narrowband case \citep{purify}, resolution of (\ref{linsystemX}) in step I. 
is the  bottleneck of the algorithm. 
Note that the possibility to solve the  $l_\nu$ systems in parallel does not increase the computation
time of this step compared to the narrowband case.

As in step V., computation of the second term of 
$\bQ_\ell$ defined in Eq. (\ref{matQ})  simplifies when $\bW_s$ is a union of $n_s$ orthonormal bases:
\begin{equation}\label{Qlsimp}
\bQ_\ell = \bH_\ell^\top\bH_\ell + \rho\bI,\quad \rho = \mu_\epsilon +\rho_S+\rho_P+n_s\rho_T
\end{equation}
The  resolution of each linear system can then be drastically
accelerated using a conjugate gradient algorithm \citep{HestenesStiefel}.
However, it is worth noting  that in the dirty image model (\ref{nbmodel}) $\bH_\ell$ is a convolution
matrix and consequently $\bQ_\ell$ in (\ref{Qlsimp}) is also a convolution matrix. Therefore  $\bQ_\ell^{-1}\bb_\ell$ can
be efficiently computed by the convolution of $\bb_\ell$ with the filter with frequency response
$(|\hat{h}_\ell(m,n)|^2 +\rho)^{-1}$ where $\hat{h}_\ell$ is the Fourier transform of the 
point spread function associated to $\bH_\ell$.
Note the obvious role of $\rho$ as a regularizer of this inversion when expressed in the frequency domain.

\section{Simulations}

A Julia implementation of the code is available\footnote{ 
\texttt{https://github.com/andferrari/muffin.jl}}.  
It takes the $\ell_\nu$ PSFs and dirty images for entries. 
$\bW_s$ is a dictionary composed by the concatenation of the first eight orthonormal 
Daubechies wavelet bases (Db1-Db8) as in \citep{purify}  and a Haar wavelet basis. 
In order to promote smoothness $\bW_\nu$ implements a Discrete Cosine Transform (DCT).
As  in \citep{Wenger:2014ty}, Chebyshev polynomial basis functions should 
be also appropriate. The implementation takes advantage of the Julia language
parallel processing support.

\begin{figure}
\includegraphics[width=.7\textwidth]{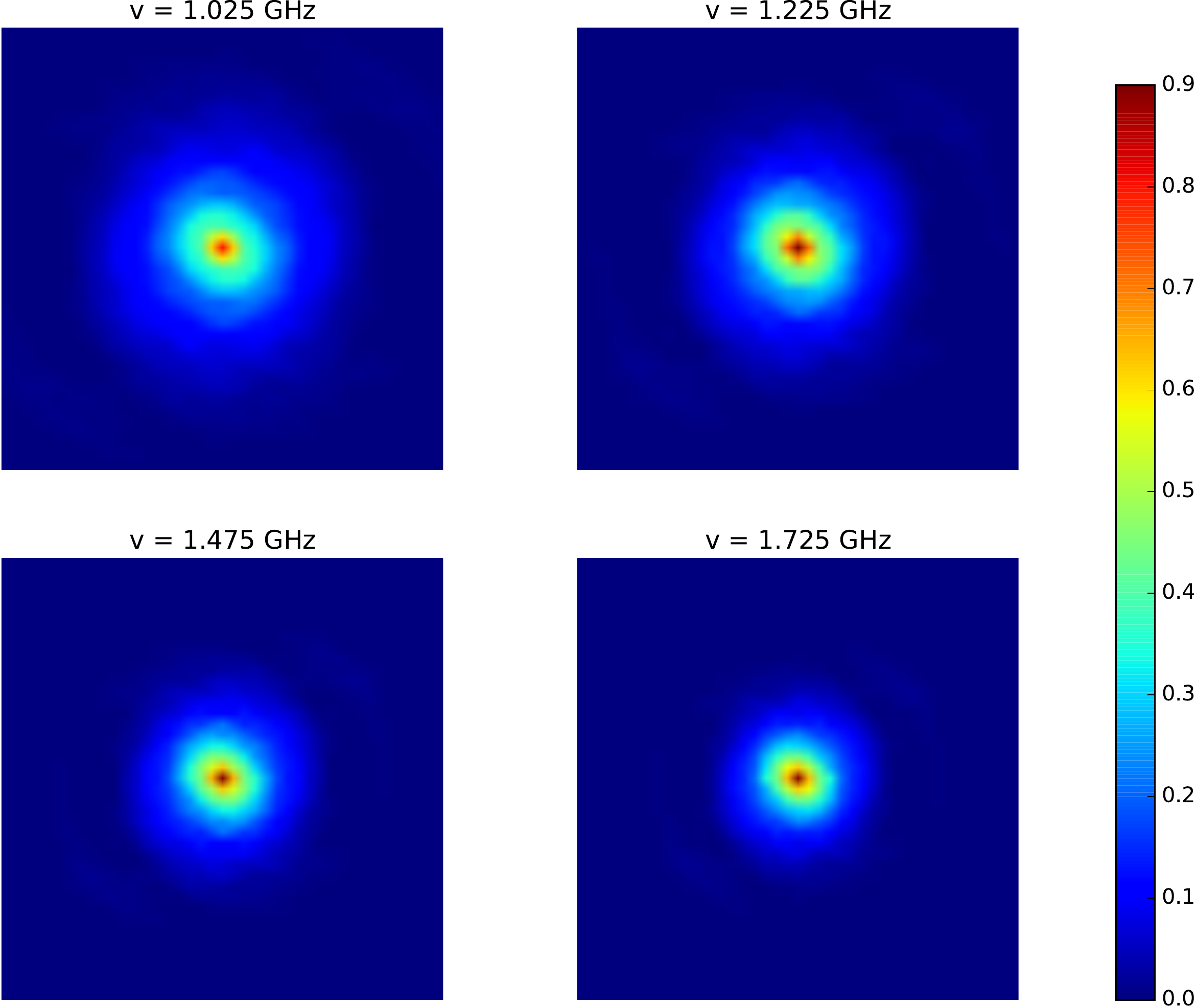}
\caption{PSFs of MeerKAT in 4 among the 15 simulated frequency  bands. 
The PSFs are obtained from MeqTrees.
The total observation time is 8 hours. 
The images are a $50\times 50$ pixels zoom of the central region. \label{fig:psf}}
\end{figure}

The PSFs used in this
simulation where computed using the MeqTrees package \citep{meqtrees} with
MeerKAT arrays configuration. We simulated $\ell_\nu=15$ frequency bands. The first is centered 
at 1.025 GHz. The central frequencies of the 15 bands are separated by 50~MHz. 
The total observation time is 8 hours. The size in pixels of the images are $256\times 256$.
Figure \ref{fig:psf} shows $50\times 50$ pixels images of the center of the PSF at 
four  frequencies (1.025 HZ, 1.225 GHz, 1.475 GHz and 1.725GHz).
The parameters $\mu_\epsilon$,  $\mu_s$ and $\mu_\nu$ control the levels 
of regularization (resp. smoothing, spatial and spectral). 
They have been set to  $\mu_\epsilon = 0.001$, $\mu_s = 0.5$ and $\mu_\nu=1.0$.
The augmented Lagragian parameters $\rho_P$, $\rho_T$,  $\rho_V$ and
$\rho_S$ affect the speed of convergence. They have been set to $\rho_P = 1.0$, 
$\rho_T = 5.0$, $\rho_V = 2.0$ and $\rho_S = 1.0$.
The classical termination criterion for the ADMM algorithm is that the primal
and dual residuals must be small \citep{Boyd10}.
In order to avoid additional parameters, 
the algorithm will be simply stopped after 350 iterations in the following simulations.

The first simulation evaluates the ability of the algorithm to reconstruct 
spectra. The sky image  is similar to the first simulation of 
\citep{RauMSMF}. It consists in two overlapping Gaussian profiles 
centered at pixel 108 and 148, with  constant spectral
indices equal  to respectively -1.0 and +1.0.

\begin{figure}
\includegraphics[angle=90,origin=c,width=.8\textwidth]{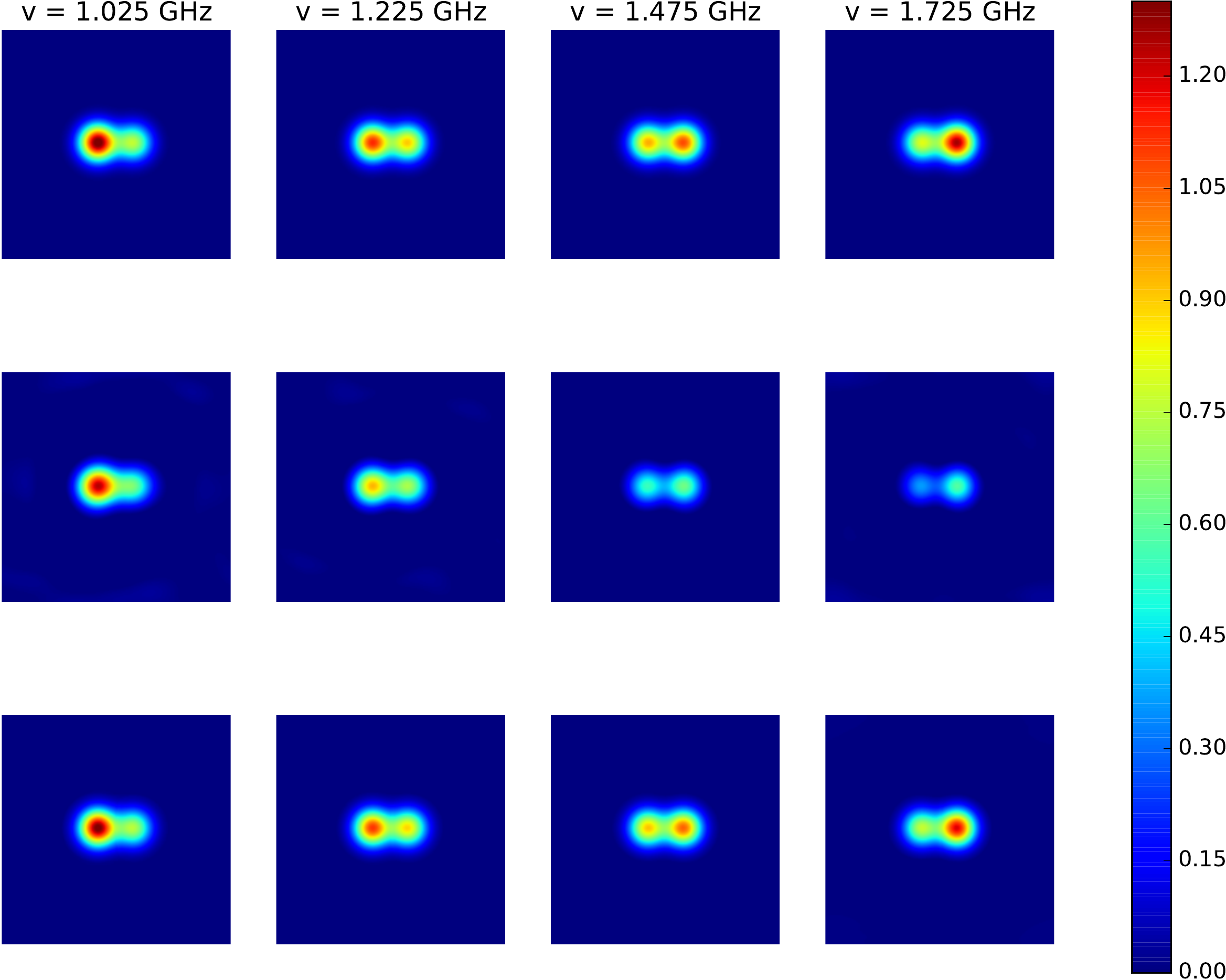}
\caption{Multi-frequency reconstruction results. 
Each line is associated to a frequency band.
The first column is the original object.  The second column is the
dirty image and the third column the reconstructed object. 
The second column (dirty images) is scaled down 
in order to fit to the color map of the two other columns.
\label{fig:reggauss}}
\end{figure}

\begin{figure}
\includegraphics[width=.7\textwidth]{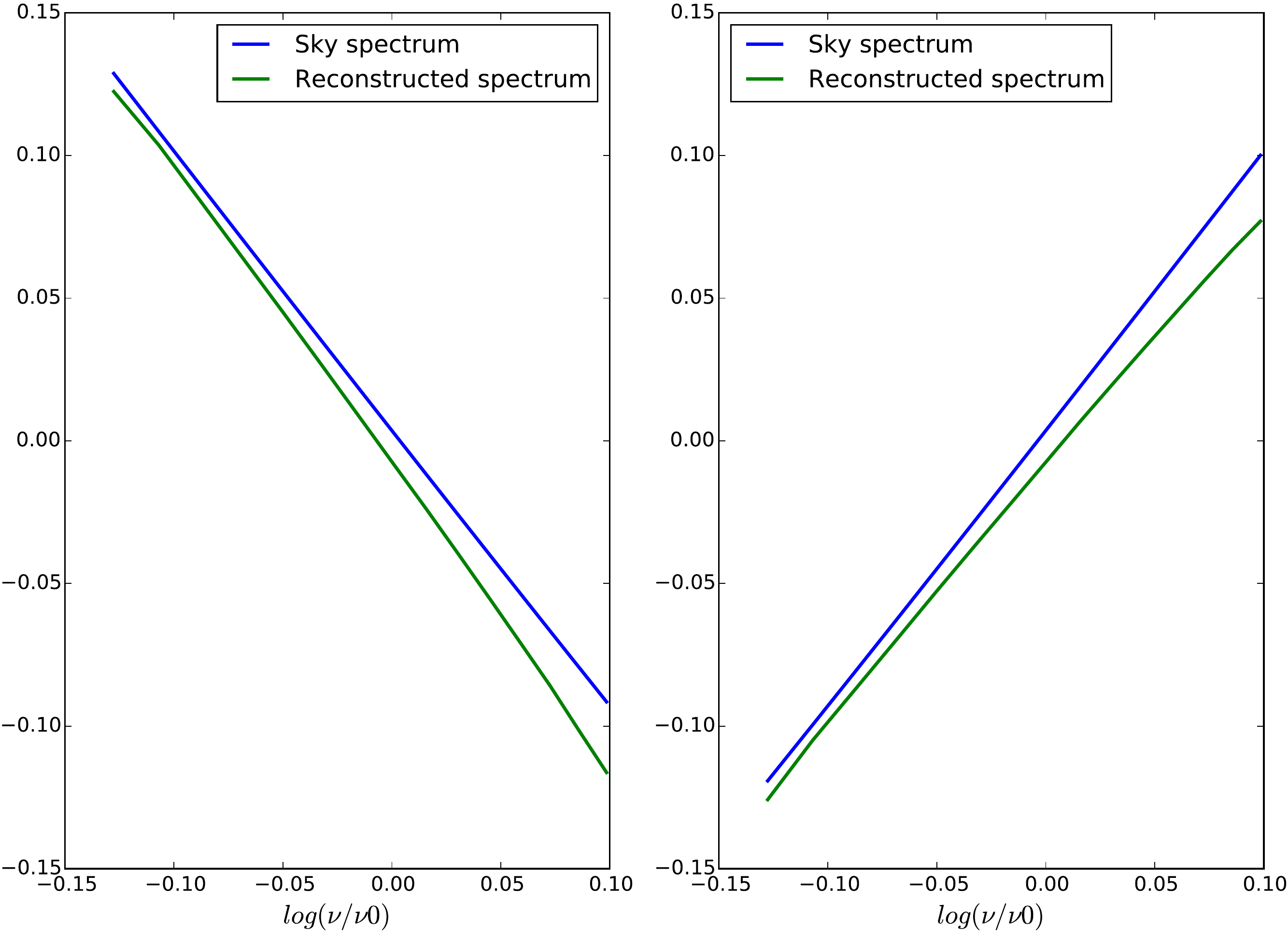}
\caption{Spectra associated to the pixel at the center of each Gaussian profile. 
Sky model in blue and reconstructed sky model in green.
\label{fig:spectrum}}
\end{figure}

\begin{figure}
\includegraphics[width=.6\textwidth]{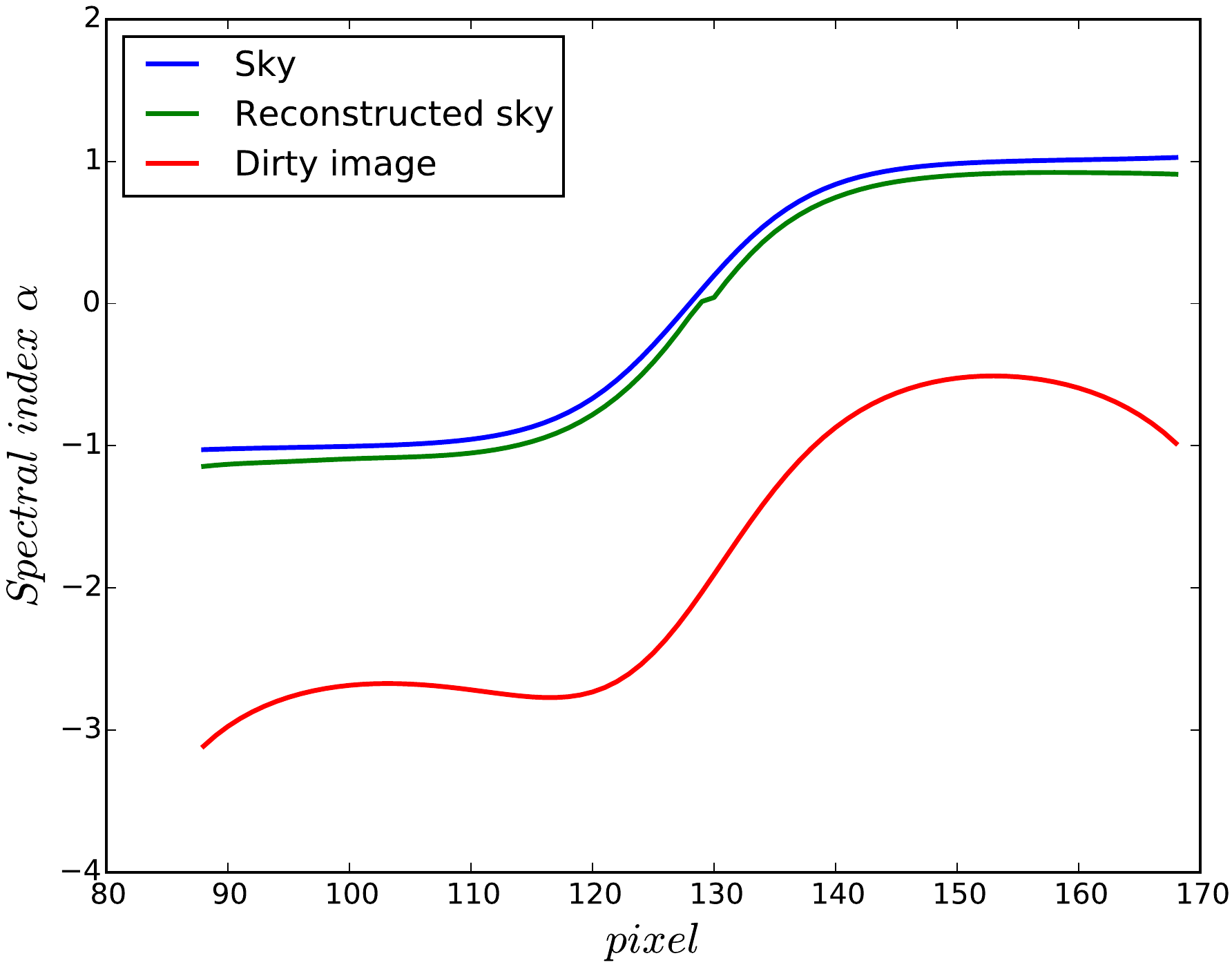}
\caption{Estimated spectral indices for the pixels on the line
joining the center of the two Gaussian profiles. Sky model in blue, reconstructed
sky model in green  and dirty image in red. \label{fig:alpha}}
\end{figure}

Figure \ref{fig:reggauss} shows in four frequency bands:
the sky image, the dirty image and the reconstructed sky obtained using 
algorithm \ref{algo:ADMMalgo}. Figure \ref{fig:spectrum}  shows the spectra
associated to the pixel  at the center of each Gaussian.
In each case the theoretical spectrum
and the reconstructed spectrum are plotted.
In order to derive a synchrotron spectral index map, each spectrum of the 
original sky (which equals here a linear combination of $\nu$ and $1/\nu$) 
is fitted to a second order power law model; i.e.
$\alpha\log(\nu)+\beta\log^2(\nu)$ in a log scale.
Spectral indices are then estimated fitting the
same model to the  reconstructed spectra.
Figure \ref{fig:alpha} shows the estimated $\alpha$ for the pixels on the line
joining the center of the two Gaussian profiles. The spectral
indices estimated directly from the dirty image are given in
the figure for comparison.
These preliminary simulations show the ability of the proposed method to
reconstruct complex spectral signatures using a non parametric approach.

\begin{figure}
\includegraphics[width=.6\textwidth]{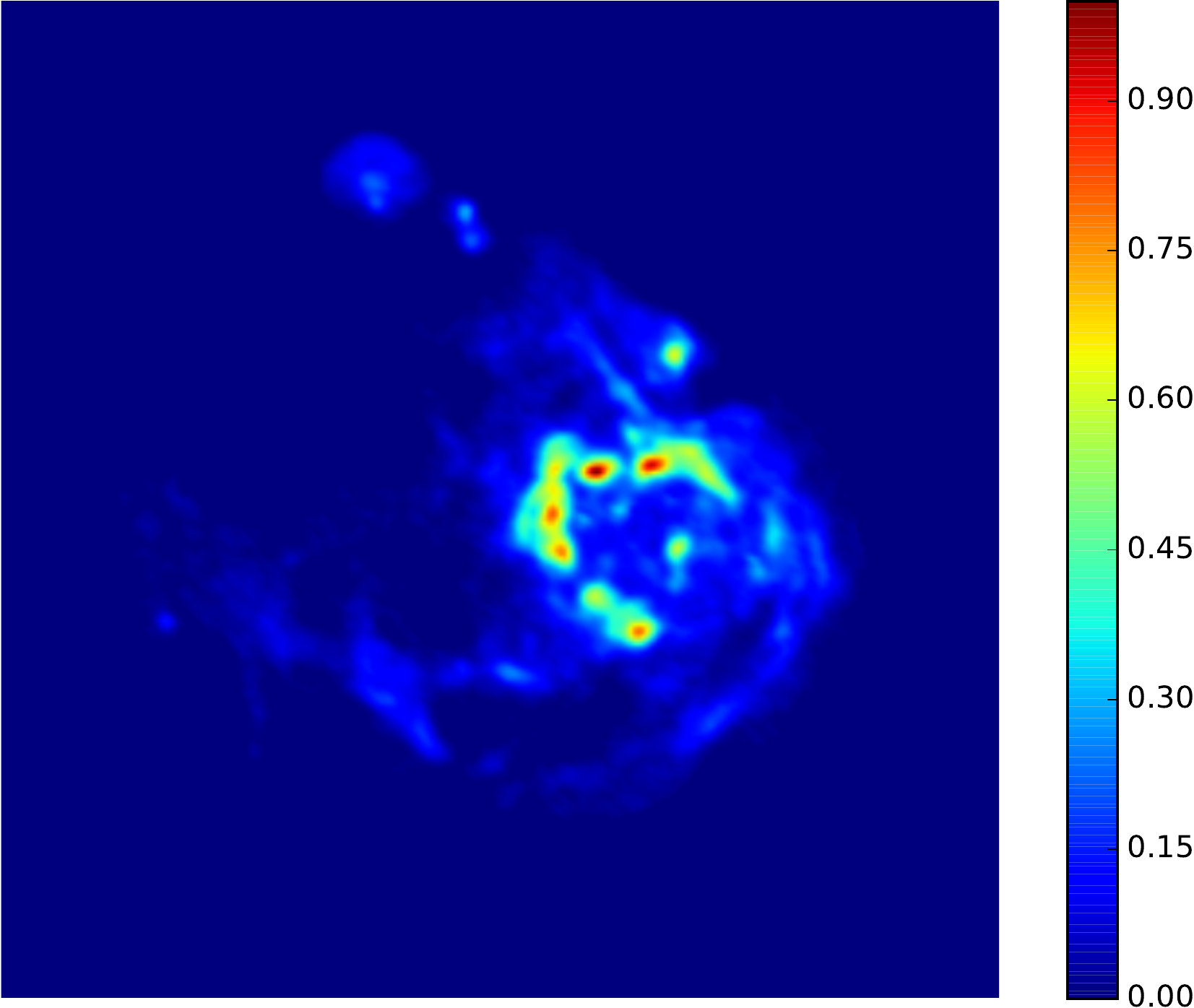}\\
\includegraphics[width=.6\textwidth]{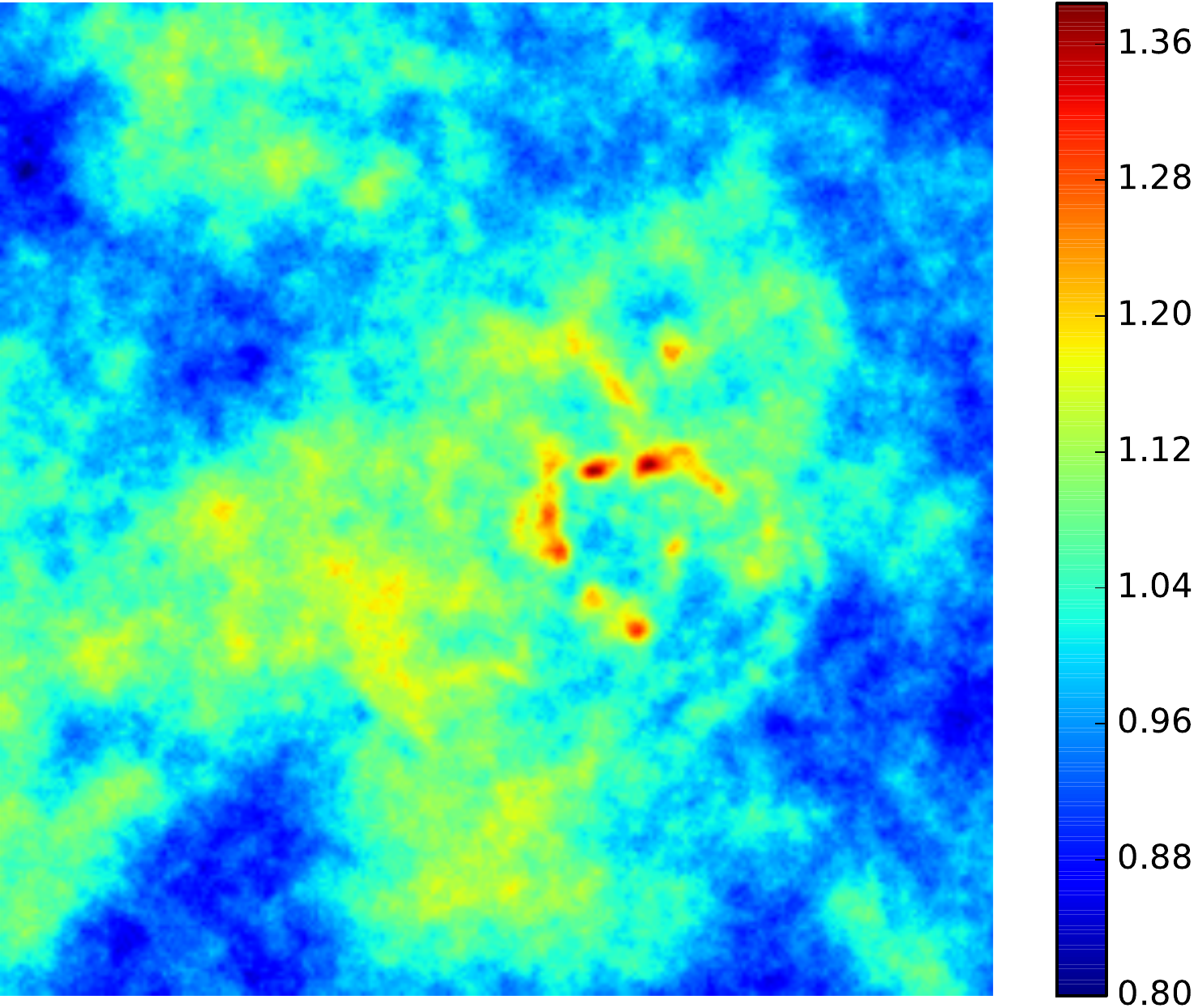}
\caption{Radio emission from an HII region in the M31 galaxy and a 
spectral indices map generated combining
the sky intensity and a Gaussian homogeneous random field. \label{fig:alphasky}}
\end{figure}

The scope of the second simulation is to evaluate the performances of the method
in a more realistic scenario. 
Radio emission from an HII region in the M31 galaxy is used as a reference sky image.
A sky cube is then computed applying a first order power-law spectrum model
to the M31 image.  
The $256\times 256$ map of spectral indices is constructed following the
procedure detailed in \citep{Junklewitz14}: for each pixel
the spectral index $\alpha$ is a linear combination of an homogeneous Gaussian field and
the reference sky image. The first correlates the spectral indices spatially and
the second correlates in a pixel the spectral index with the object intensity.
Figure \ref{fig:alphasky} shows the sky map and a typical spectral indices map.
A white gaussian noise with a constant variance corresponding to a
signal to noise ratio of 20 dB has been added to all the dirty images.

\begin{figure}
\includegraphics[angle=90,origin=c,width=.8\textwidth]{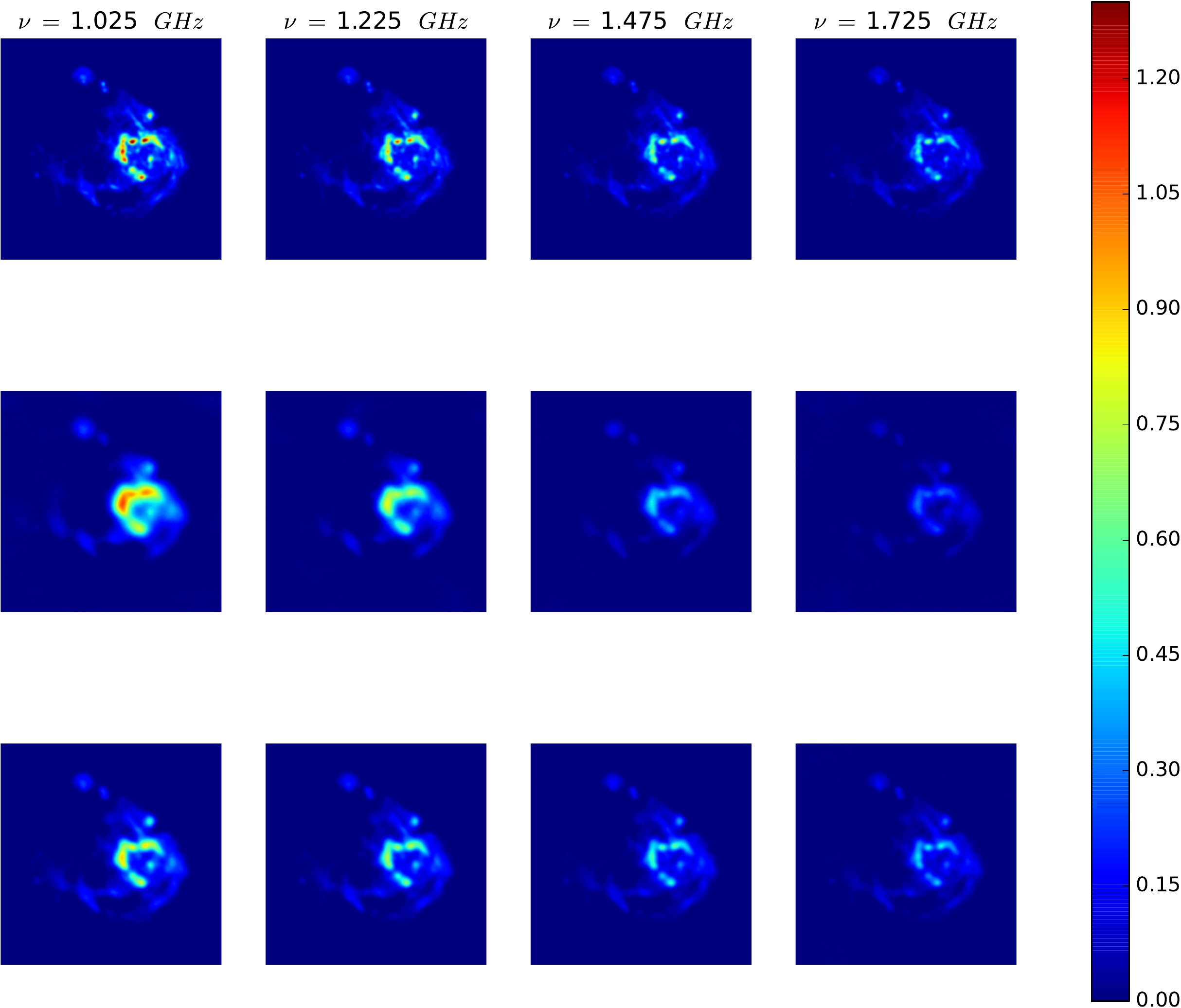}
\caption{
Multi-frequency reconstruction results. Each line is associated to a frequency band. The first column is the original sky. The second column is the dirty image and the third column  the reconstructed sky. The second column (dirty images) is scaled down 
in order to fit to the color map of the two other columns. \label{fig:m31deconv}}
\end{figure}

\begin{figure}
\includegraphics[width=.6\textwidth]{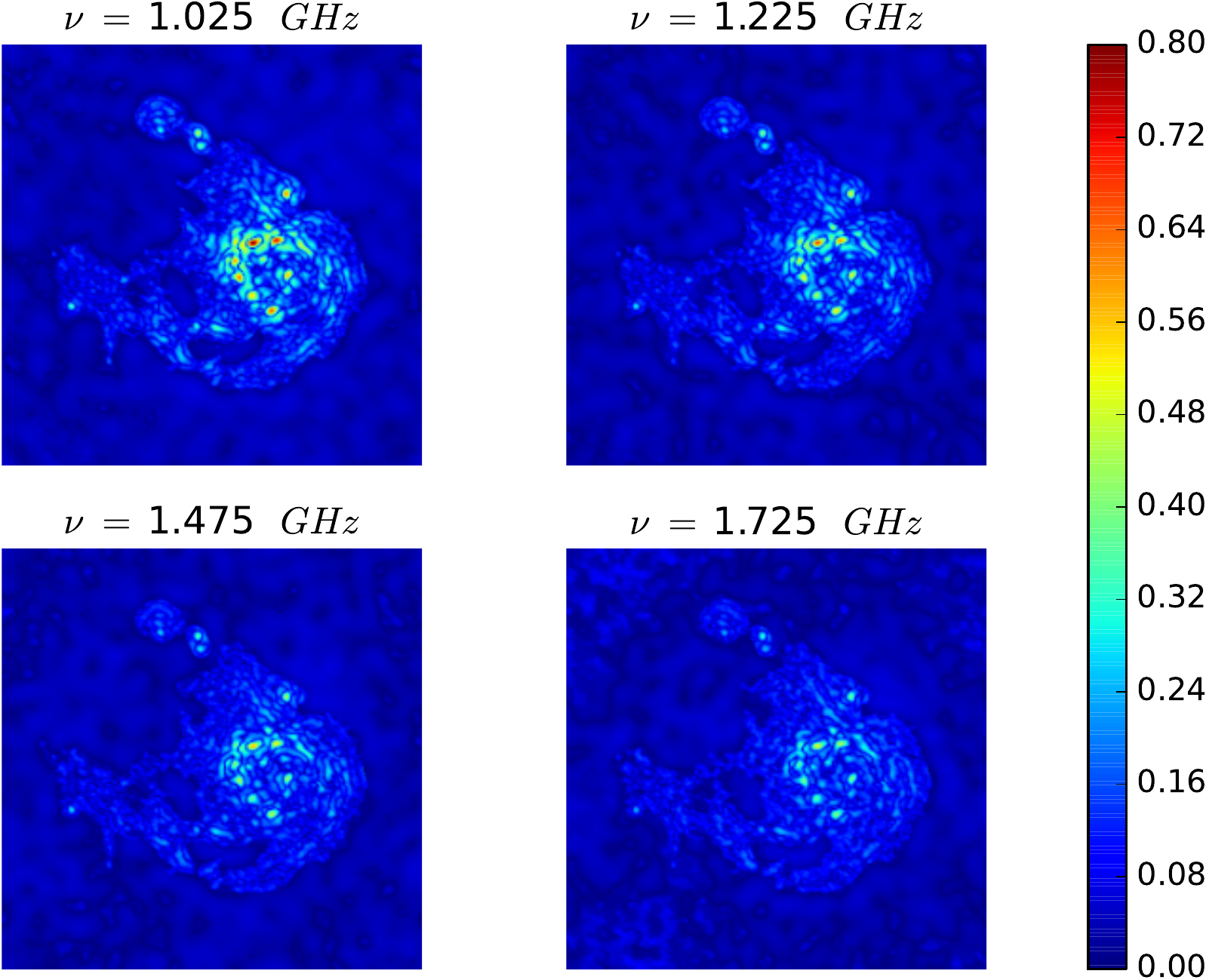}
\caption{
Error between the sky and the reconstructed sky. The
images show the square root of the absolute value of the difference
between the two images.\label{fig:errm31}}
\end{figure}

\begin{figure}
\includegraphics[width=.6\textwidth]{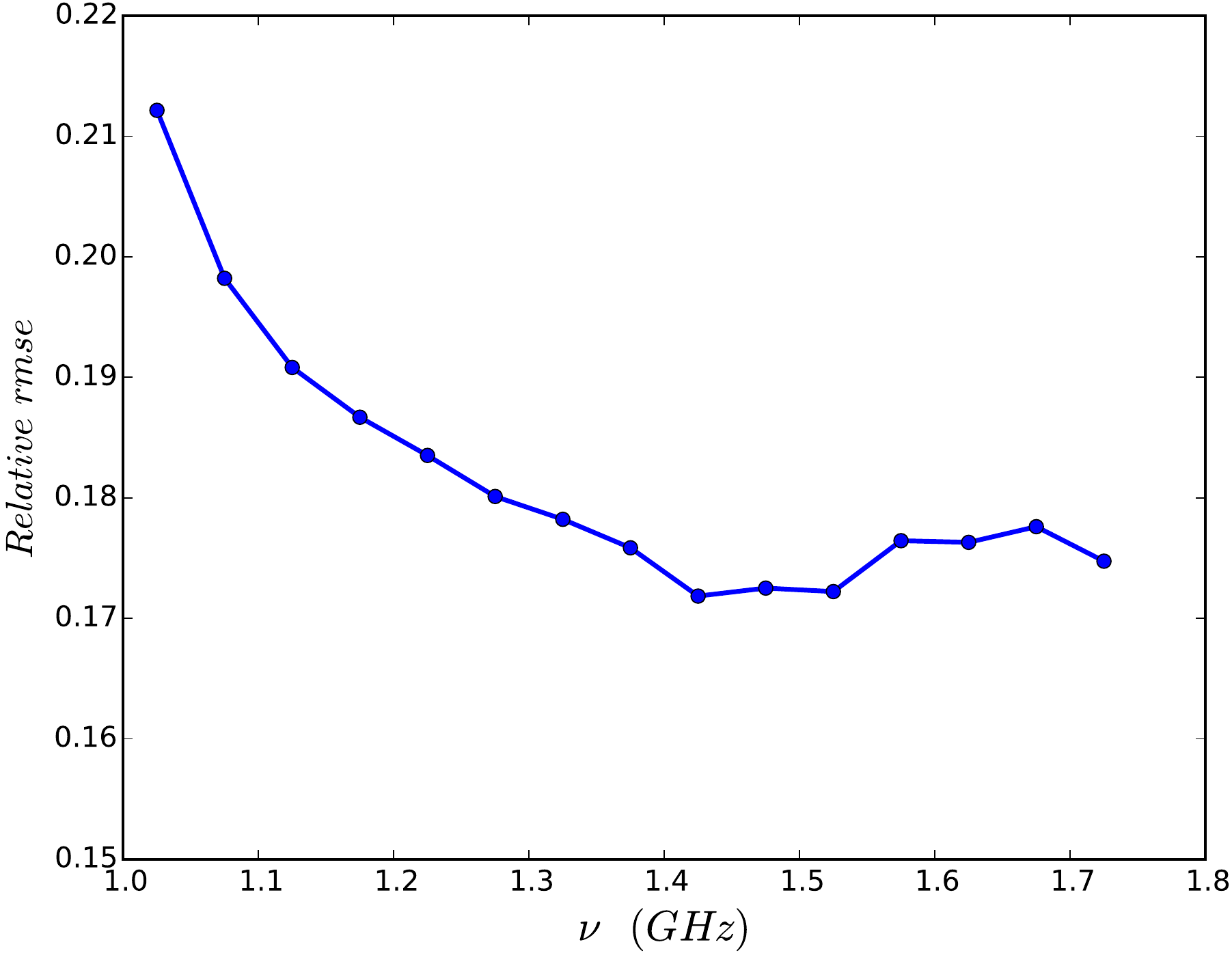}
\caption{Relative root mean square error (RMSE) of the reconstructed sky as a function
of the frequency. \label{fig:rrmse}}
\end{figure}

The computation time required to reconstruct the $256\times 256\times 15$ pixels cube,
using $\nu_\ell=15$ cores\footnote{\texttt{http://calculs.unice.fr}}
is approximately 1.5 hours. Figure \ref{fig:m31deconv} shows in four frequency bands:
the sky image, the dirty image and the reconstructed map.
Figure \ref{fig:errm31} shows for the same frequency bands
the error images between the sky and the reconstructed sky
(square root of the absolute value of the difference
between the two images).
Finally, figure \ref{fig:rrmse} gives the relative root mean square error (RMSE) as a function
of the frequency band. These results clearly show the capacity of the algorithm 
to recover details of the sky image at every frequency.

\section{Conclusion}

This paper presents first results on multi-frequency image 
reconstruction for radio interferometry using a fully regularized 
inverse problem approach.
The proposed algorithm relies on the minimisation of a data fidelity term 
spatially and spectrally regularized. An efficient iterative algorithm is derived
to minimize the related convex cost function. The computational bottleneck of the algorithm can be parallelized w.r.t. the images at each frequency leading to an overall computation
time of the order of  narrow-band algorithms. Future work will focus among others
on the derivation of specific spectral regularizations and on the necessity of taking
into account frequency-dependent instrumental effects such as a the primary-beam.

\clearpage

\bibliographystyle{plainnat}
\bibliography{bibli}

\end{document}